\documentclass{skads2009}

\usepackage{graphicx}

\begin{document}

\title{Exploring the magnetic fields in local and distant galaxies}

   \titlerunning{Exploring the magnetic fields in local and distant galaxies}

   \authorrunning{Arshakian et al.}

   \author{T.\,G. Arshakian\inst{1}, R. Stepanov \inst{2}, R. Beck \inst{1}, M. Krause \inst{1}, D. Sokoloff    \inst{3}
    \and P. Frick \inst{2} }

%   \offprints{T.\,G. Arshakian}

   \institute{Max-Planck-Institut f\"ur Radioastronomie, Auf dem H\"ugel 69,
   53121 Bonn, Germany \thanks{This work was supported by the
   European Commission Framework Program 6, Project SKADS, Square
   Kilometre Array Design Studies (SKADS), contract no 011938.}\\
   \email{tigar@mpifr-bonn.mpg.de}
         \and
  Institute of Continuous Media Mechanics, Korolyov str.~1,
  614061 Perm, Russia
         \and
  Department of Physics, Moscow State University, Moscow 19899, Russia\\
             }

  \abstract{
The dominant
population of sources at low flux densities observable with future
radio telescopes is the population of star-forming disk galaxies
which, e.g. with the forthcoming Square Kilometre Array (SKA), would
be detected up to about $z\sim3$ in total radio intensity, and up to
$z\sim5$ with Faraday rotation measures (RM) of intervening disk
galaxies towards polarized background quasars. We investigate the possibility to recognize the magnetic field
structures in nearby galaxies and to test the cosmological evolution
of their large- and small-scale magnetic fields with the SKA and its
precursors. We estimate the required density of the background polarized sources
detected with the SKA for reliable recognition and reconstruction
of magnetic field structures in nearby spiral galaxies. The dynamo
theory is applied to distant galaxies to explore the evolution of
magnetic fields in distant galaxies in the context of a hierarchical
dark matter cosmology. Under favorite conditions, a \emph{recognition} of large-scale
magnetic structures in local star-forming disk galaxies (at a
distance $\la 100$ Mpc) is possible from $\ga 10$ RMs towards
background polarized sources. Galaxies with strong turbulence or
small inclination need more polarized sources for a statistically
reliable recognition. A reliable \emph{reconstruction} of the field
structure without precognition needs at least 20 RM values on a cut
along the projected minor axis which translates to $\approx1200$
sources towards the galaxy. We demonstrate that early regular fields
are already in place at $z \sim 3$ (approximately 1.5 Gyr after the
disk formation) in massive gas-rich galaxies ($>10^9$ M$_{\sun}$)
which then evolve to Milky-Way type galaxies. Major and minor
mergers influence the star formation rate and geometry of the disk
which has an effect of shifting the generation of regular fields in
disks to later epochs. Predictions of the evolutionary model of
regular fields, simulations of the evolution of turbulent and
large-scale regular fields, total and polarized radio emission of
disk galaxies, as well as future observational tests with the SKA
are discussed. }
  \maketitle
%
%________________________________________________________________

\section{Introduction}
Most of what we know about galactic magnetic fields comes through
the detection of radio waves. {\em Synchrotron emission} is related
to the total field strength in the sky plane, while its polarization
yields the orientation of the regular field in the sky plane and
also gives the field's degree of ordering. Incorporating {\em
Faraday rotation} provides information on the strength and direction
of the coherent field component along the line of sight. Faraday
rotation measures (RM) in galaxies are generated by regular fields
of the galaxy plus its ionized gas, both of which extend to large
galactic radii. RM towards polarized background sources can trace
regular magnetic fields in these galaxies out to even larger
distances, however, with the sensitivity of present-day radio
telescopes, the number density of polarized background sources at
1.4\,GHz is only a few sources per solid angle of a square degree,
so that only M~31 and the LMC could be investigated so far 
(Han et al. \cite{han98}, Gaensler et al. \cite{gaensler05}).

The amplitude and structure of magnetic fields in local galaxies is
successfully reproduced by the mean-field dynamo theory
(Beck et al. \cite{beck96}). This suggests that the dynamo theory can also be
applied for distant galaxies to explore the evolution of magnetic
fields at high redshifts. The evolution of magnetic fields in
galaxies is coupled to the formation and evolution of disk galaxies
which are fundamental problems in astronomy. Recent high-resolution
numerical simulations of disk formation in galaxies showed that a
stable disk could be formed at redshifts $z\sim 5-6$ and even higher
(Governato et al. \cite{governato04}, Mayer et al. \cite{mayer08}). A better understanding of the history
of magnetism in young galaxies may help to solve fundamental
cosmological questions on the formation and evolution of galaxies
(Gaensler et al. \cite{gaensler04}, Arshakian et al. \cite{arshakian09}).

Future high-sensitivity radio facilities will observe polarized
intensity and RM for a huge number of faint radio sources, thus
providing a high density background of polarized point sources. This
opens the possibility to study in detail the large-scale patterns of
magnetic fields and their superpositions, thus allowing tests of the
dynamo theory for field amplification and its ordering. A major step
towards a better understanding of galactic magnetism will be
achieved by the SKA (www.skatelescope.org) and its precursors
(ASKAP, MeerKAT).

%Throughout the paper a flat cosmology model is used with
%$\Omega_{m}=0.3$ ($\Omega_{\Lambda}+\Omega_{m}=1$) and $H_0=70$
%km\,s$^{-1}$\,Mpc$^{-1}$.
\begin{figure}[t]
\begin{center}
\includegraphics[width=0.49\textwidth]{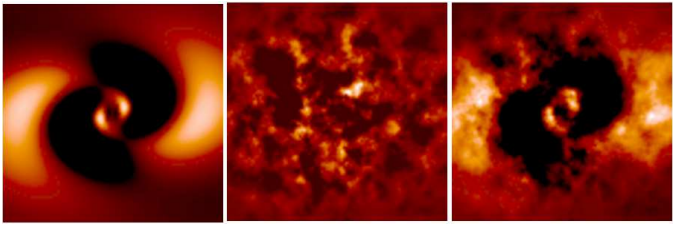}
\caption {\small\emph{Left panel}: RM maps (the scale is from +150 to -150 rad m$^{-2}$)
generated for a typical spiral galaxy of radius 10\,kpc (the frame size is $20\times20$ kpc) with a pure
bisymmetric magnetic field structure (regular magnetic field
strength of 5~$\mu$G) and thermal electron density
$n_0=0.03$~cm$^{-3}$, at the inclination angle of $i=10^o$.
\emph{Middle panel}: Modeled random turbulent field for
30~rad~m$^{-2}$ r.m.s. \emph{Right panel}: In the right panel is the
`realistic' map of RM of the galaxy obtained by superposition of
these two field components.} \label{fig:rmr}
\end{center}
\end{figure}

\section{Recognition of magnetic field structures in local disk galaxies}

\label{sec:local} The presence of regular kiloparsec-scale magnetic
fields in nearby spiral galaxies makes Faraday rotation an ideal
tool for studying the structure of magnetic fields in the disk and
halo. With the SKA capabilities RMs of hundred and thousands
polarized background sources can be measured behind nearby galaxies
thus allowing their detailed magnetic field mapping. We estimated
the required density of the background polarized sources for
reliable recognition and reconstruction of the magnetic field
structure in nearby spiral galaxies (Stepanov et al. \cite{stepanov08}). For a
typical spiral galaxy we modeled the distribution of the ionized gas
and the magnetic field in the disk (axisymmetric, bisymmetric and
quadrisymmetric spiral, and superpositions). We simulated the RM of
a galaxy towards background polarized sources taking into account
the RM fluctuations as result of turbulent fields and/or
fluctuations in ionized gas (Fig.~\ref{fig:rmr}). The simulated
magnitudes of RM towards background point sources and their density
are then used to recognize the magnetic field structure and assess
its reliability.

The slope of the number counts ($\gamma$) of polarized background
sources is unknown beyond 0.5~mJy at 1.4~GHz and important for the
field recognition. For the optimistic value of $\gamma=-1.1$, the
\emph{recognition} of a single and mixed modes of regular fields can
be reliably performed from a limited sample ($\ga 10$) of RM
measurements of polarized point sources. Single higher modes
(bisymmetric and quadrisymmetric spirals) are easier to recognize,
i.e. they need less RM points, shorter observation time and are less
affected by the turbulent component. The dependence on turbulence
becomes dramatic for weakly inclined (almost face-on) galaxies --
reliable fitting requires a huge number of sources. Future deep
(100~h) RM observations with the SKA at 1.4~GHz will allow the
recognition of $\la 60000$ spiral galaxies in the volume restricted
to a radius of $\sim 100$~Mpc.

The \emph{reconstruction} of magnetic field structures of strongly
inclined galaxies without precognition of a simple pattern is
possible for a large sample ($>1000$) of RM point sources. This
would require a sensitivity of the SKA at 1.4~GHz of $\approx
0.5-5~\mu$Jy (or integration time less than one hour) for galaxies
at distances of about 1~Mpc. The field structures of $\sim50$
galaxies until about 10~Mpc distance can be reconstructed with tens
to hundred hours of integration time.

Radio telescopes operating at low frequencies (LOFAR, ASKAP, and the
low-frequency SKA array) may also be useful instruments for field
recognition or reconstruction with the help of RM, if background
sources are still significantly polarized at low frequencies. The later reduces the
number of RM points per solid angle of the galaxy: (i) the
overall degree of polarization is most probably lower due to Faraday
depolarization effects, (ii) the typical degree of polarization
may decrease for more distant sources (stronger star formation in young galaxies), and (iii) internal and external depolarization may significantly reduce
the number of polarized sources. Another problem 
at low frequencies is thermal absorption that may
reduce the radio fluxes at frequencies below about 100~MHz for
strongly inclined galaxies with high densities of the ionized gas.
For planning surveys at frequencies of 300~MHz and lower, much more needs
to be known about the statistics of polarized sources.

\section{Three-phase evolutionary model of magnetic fields in galaxies}

\label{sec:distant} 
\begin{figure}
  \begin{center}
    \includegraphics[width=0.45\textwidth]{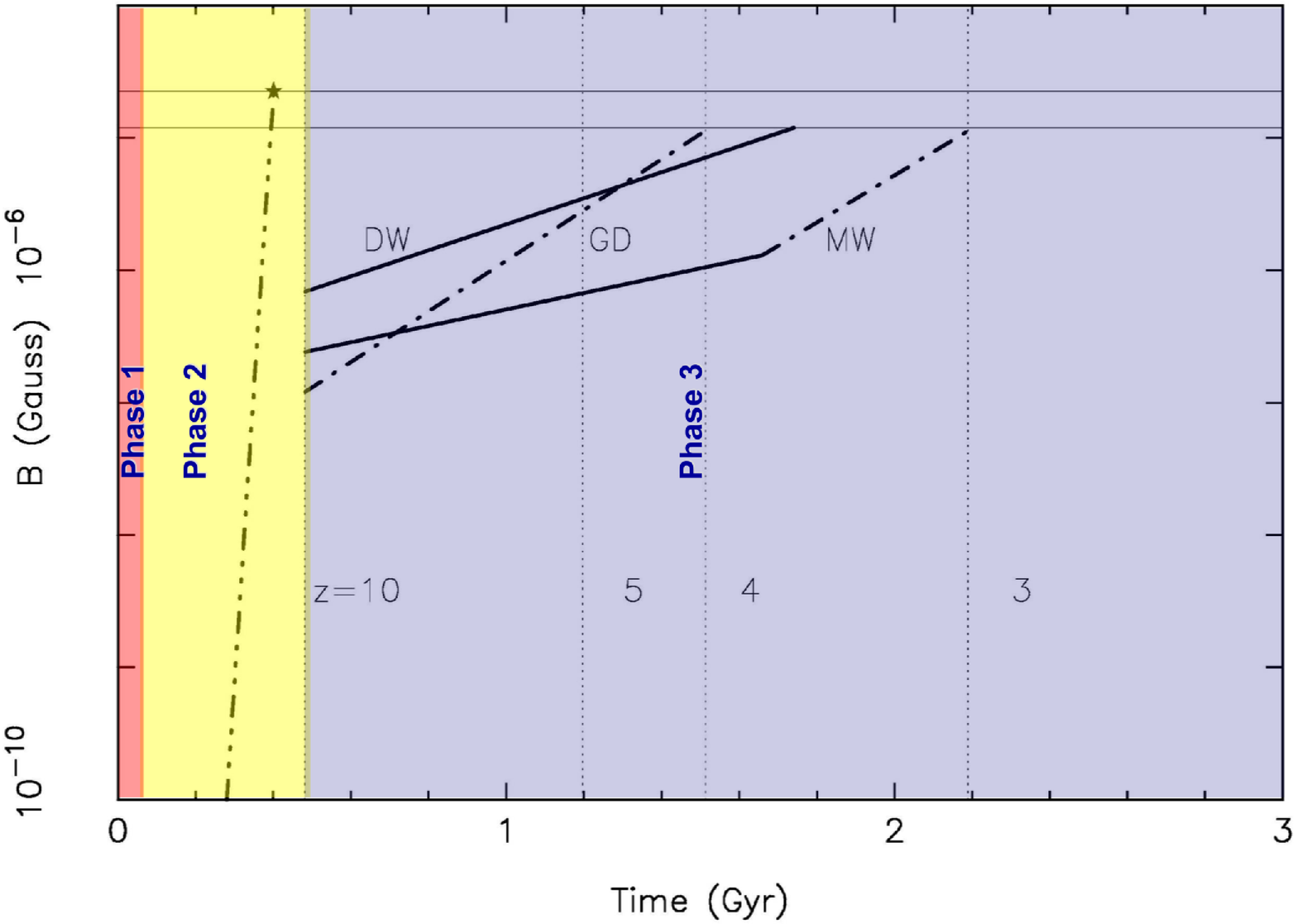}
    \includegraphics[width=0.45\textwidth]{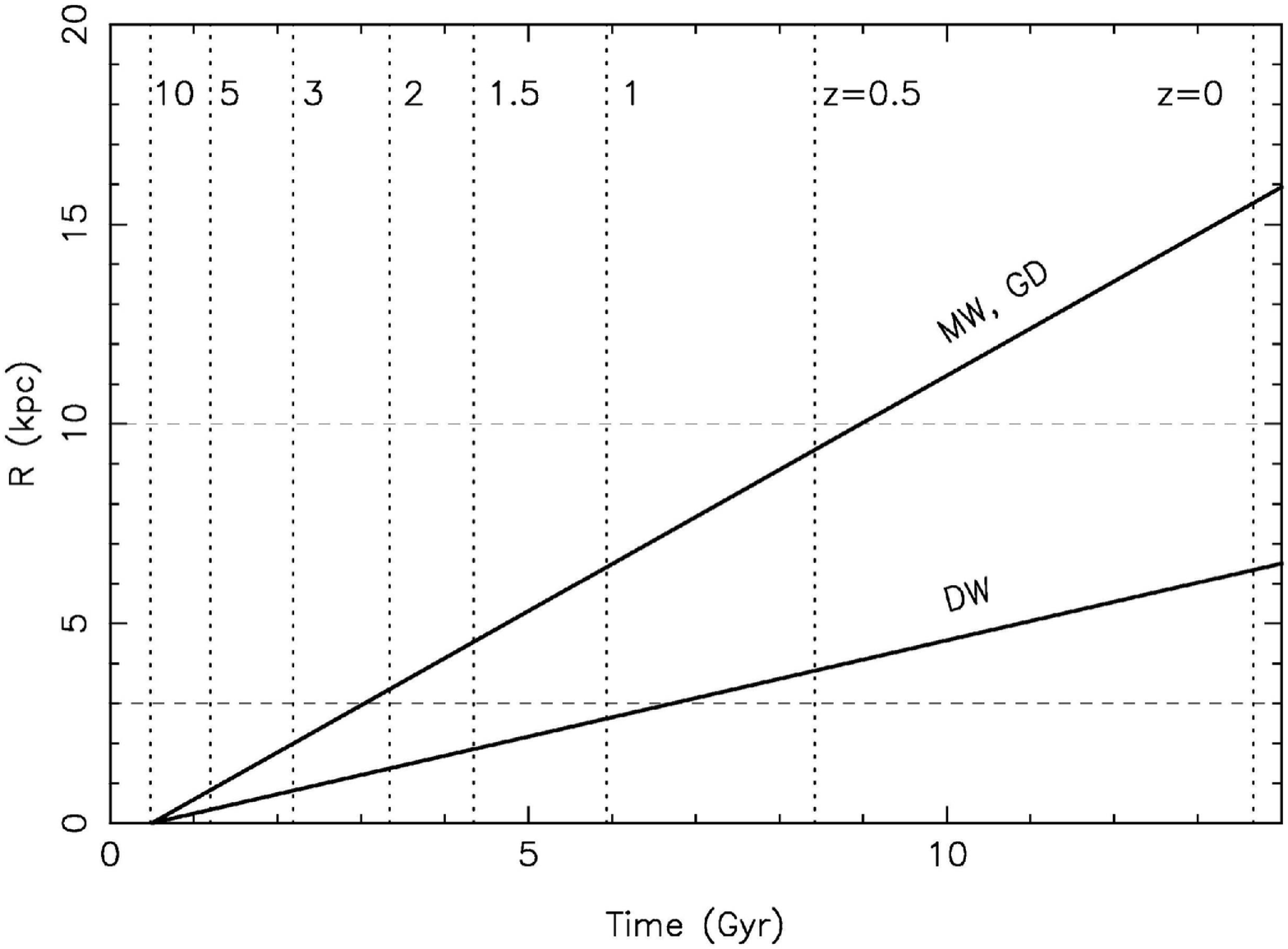}
  \end{center}
  \caption{\small{\emph{Top panel}: Evolution of magnetic field
  strength in dwarf galaxies (DW), MW-type galaxies and in giant disk (GD)
  galaxies. Phase 1 (red): Origin of seed magnetic fields.
  Phase 2 (yellow): Evolution of the small-scale magnetic field
  generated by the turbulent dynamo (thick dashed-dot-dot line).
  Phase 3 (blue): Evolution of the large-scale magnetic field
  generated by the mean-field dynamo in quasi-spherical galaxies
  (thick solid line) or in thin-disk galaxies (thick dashed-dot-dashed).
  At $z\approx4$ the dynamo type in Milky-Way type galaxies switched
  from spherical to thin-disk. \emph{Bottom panel}: Evolution of the ordering scale of regular
  magnetic fields for dwarf galaxies (DW; bottom line) and disk galaxies (MW and GD; top line).}}
  \label{fig:mfe}
\end{figure}

Studying the evolution in magnetic fields of
galaxies is important for interpreting future radio synchrotron
observations with the SKA. We have used the dynamo theory to derive
the timescales of amplification and ordering of magnetic fields in
disk and quasi-spherical galaxies (Arshakian et al. \cite{arshakian09}). This has
provided a useful tool in developing a three-phase evolutionary
model of regular magnetic fields, coupled with models describing the
formation and evolution of galaxies. In the hierarchical structure
formation scenario, we identified three main phases of
magnetic-field evolution in galaxies (Fig.~\ref{fig:mfe}). In the
epoch of \emph{dark matter halo formation} (first phase), seed
magnetic fields of $\sim10^{-18}$~G strength were generated in
protogalaxies by the Biermann battery or the Weibel instability
\cite{lazar09}. Turbulence in the protogalactic halo generated by
thermal virialization could have driven the turbulent (small-scale)
dynamo and amplified the seed field to the equipartition level of
$\approx 20~\mu$G within a few $10^8$ yr (second phase). In the
epoch of \emph{disk formation}, the turbulent field served as a seed
for the mean-field (large-scale) dynamo developed in the disk (third
phase).

\begin{itemize}

\item We defined three characteristic timescales for the evolution
of galactic magnetic fields: one for the amplification of the seed
field, the second one for the amplification of the large-scale
regular field, and the third one for the coherent field ordering on
the galactic scale (Arshakian et al. \cite{arshakian09}).

\item Galaxies similar to the Milky Way (MW) formed a thick disk at
$z\approx10$, and the mean-field spherical dynamo amplified the field until
$z\approx4$, at which the disk became sufficiently thin, so that the more
efficient mean-field disk dynamo could operate and amplified the regular fields within 1.5~Gyr.
Regular fields of
equipartition (several $\mu$G) strength and a few kpc coherence
length were generated within 2~Gyr (until $z\approx3$), but field
ordering up to the coherence scale of the galaxy size took another
6~Gyr (until $z\approx0.5$). Giant galaxies (radius $>15$~kpc) had
already formed a thin disk at $z\approx10$, allowing an efficient
dynamo generation of equipartition regular fields (with a coherence
length of about 1~kpc) until $z\approx4$. However, the Universe is
too young for fully coherent fields to have already developed in
disks of giant galaxies. Dwarf galaxies (radius $<3$~kpc) formed
even earlier. If their rotation is ordered and sufficiently fast to
allow the action of the spherical mean-field dynamo, they should
have hosted fully coherent fields at $z\approx1$.

\item Major mergers excited starbursts with enhanced
turbulence, which in turn amplified the turbulent field, whereas the
regular field was disrupted and required several Gyr to recover.
Measurement of regular fields can serve as a clock for measuring the
time since the last starburst or merger event.

\item Starbursts due to major mergers enhance the
turbulent field strength by a factor of a few and drive a fast wind
outflow, which magnetizes the intergalactic medium. Observations of
the radio emission from distant starburst galaxies can provide an
estimate of the total magnetic-field strength in the IGM.

\end{itemize}
\hspace{0.2cm}

\begin{figure}
  \begin{center}
    \includegraphics[width=0.49\textwidth]{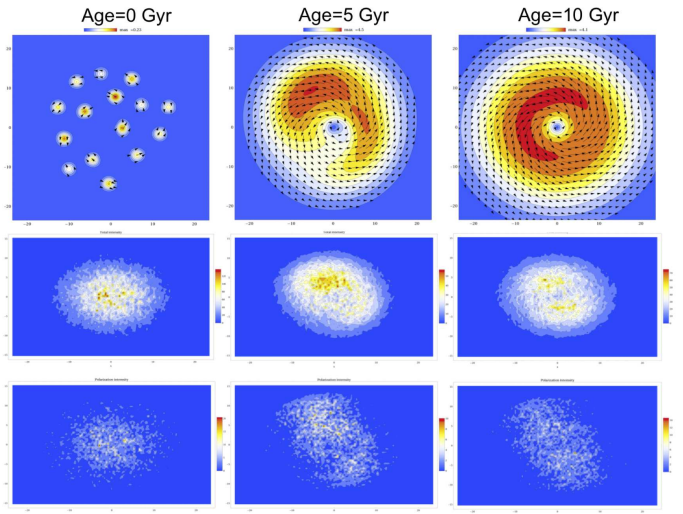}
  \end{center}

\caption{\small{Simulations (in the rest frame of a galaxy) in the framework of the SKA design
studies. \emph{First row}: the evolution of regular magnetic fields
in the disk of a galaxy seen face-on (the frame size is
20~kpc$\times$20~kpc). The amplitude and ordering scale of the
regular fields in 0.5~Gyr after the epoch of disk formation ($0.3\,\mu$G
and 1~kpc; left panel), after 5~Gyr ($\sim 1.5\,\mu$G and
6~kpc; middle panel), and after 10~Gyr ($\sim 1.5\,\mu$G and
12~kpc; right panel). Simulations of the total intensity
(\emph{second row}) and polarization (\emph{third row}) at 150~MHz
for a galaxy with an inclination angle of $60$ degrees, turbulent
(3~$\mu$G) and regular (1.5~$\mu$G) magnetic fields, and a
star-formation rate of 10 M$_{\sun}$ yr$^{-1}$ are shown for the
same epochs.
  }
  }
\label{fig:mfes}
\end{figure}

We then simulated the evolution of magnetic fields for disk galaxies
(Arshakian et al., in prep.) using the timescales of amplification
and ordering of regular magnetic fields derived in
Arshakian et al. \ (\cite{arshakian09}). For a MW-type galaxy ($R=10$~kpc), we started
from seed fields generated in 0.5 Gyr after the disk formation at $z=10$. The
size of the seed (1~kpc), amplitude of its regular field
(distributed normally around $0.3\,\mu$G) and pitch angle
(scattered within $\pm 20^{\circ}$) are used to simulate the
evolution of regular fields to later epochs (Fig.~\ref{fig:mfes}).
After 5~Gyr the regular field is amplified to the equipartition
level while the field is ordered at scales of 5~kpc. After 10~Gyr
the field is fully coherent. At low frequencies (150~MHz),
significant depolarization caused by regular fields
(Fig.~\ref{fig:mfes}; bottom panels) results in an asymmetric
polarized structure elongated near the minor axis of a galaxy.

\section{Future observational tests with the SKA}

Our analysis has important implications on the expectations of the
future observations of magnetic fields with the SKA.
%Studying the evolution in magnetic fields of galaxies is important
%for interpreting future radio synchrotron observations with the SKA.
The total magnetic field can be measured by the observed total power
radio emission, corrected for the thermal fraction of a galaxy,
while the regular magnetic field can be traced by polarized
synchrotron emission and by Faraday rotation. The tight radio --
far-infrared correlation in galaxies implies that radio synchrotron
emission is an excellent tracer of star formation in galaxies, at
least to distances of $z\simeq3$ (Seymour et al. \cite{seymour08}).

The small-scale dynamo, with the help of virial turbulence, could
have amplified turbulent fields to the level of equipartition with
turbulent energy density within $\simeq 3\times10^8$ years
(Arshakian et al. \cite{arshakian09}); strong fields should therefore existed in all
star-forming galaxies at $z\simeq10$ (Fig.~\ref{fig:mfe}; top panel)
and the radio--far-infrared correlation should be valid for
$z\la10$. However, if the total field strength was smaller than
$3.25~\mu$G~(1+z)$^2$, the strong cosmic microwave background (CMB)
radiation at high redshifts would have suppressed the non-thermal
continuum emission of a galaxy by means of inverse Compton losses of
cosmic-ray electrons, suggesting that the radio--far-infrared
correlation should evolve with infrared/radio ratios increasing with
redshift (Murphy \cite{murphy09}). Suppression of the non-thermal component
at high redshifts ($z>3$) would have left only radiation in the
thermal (free-free) regime, thus limiting the depth to which the SKA
can detect star-forming galaxies. Deviations from nominal IR/radio
ratios at high-z will provide a means for constraining the presence
and strength of magnetic fields in young galaxies (Murphy \cite{murphy09}).
The SKA and its precursor telescopes will investigate this relation
in more detail.

A number of predictions of the evolutionary model of magnetic fields
can be tested with the SKA's measurements of polarized synchrotron
emission and Faraday rotation of distant galaxies: (i) an
anticorrelation at fixed redshift between galaxy size and the ratio
between ordering scale and the galaxy size, (ii) giant galaxies
(radius $>15$~kpc) should not host fully coherent fields (no
large-scale RM patterns) until the present epoch, (iii) Milky-Way
type galaxies should host fully coherent fields (large-scale RM
patterns) at $z\la0.5$, (iv) undisturbed dwarf galaxies with ordered
rotation should host fully coherent fields giving rise to strong
Faraday rotation signals (large-scale RM patterns) at $z\la1$, 
and (v) weak regular fields (small Faraday rotation) in galaxies at
$z\la3$, possibly associated with strong anisotropic fields (strong
polarized emission), would be signatures of major mergers
(Arshakian et al. \cite{arshakian09}).

The method of observing RM against distant polarized background
sources can also be applied to measure the strength of the regular
field in distant intervening star-forming galaxies, identified by
additional optical spectroscopic observations with Mg II absorption
systems (Bernet et al. \cite{bernet08}). Deeper observations with future telescopes
are essential to provide RM data with much better statistics to larger redshifts ($z\la5$) and in
much smaller redshift bins and would allow to study the `RM
function' which we define as the number of background polarized
sources per RM and redshift intervals. This will allow the evolution
of the amplitude and ordering scale of regular fields to be tested
(Arshakian et al., in prep.).

To study the magnetic evolution of galaxies by means of RM of distant polarized sources, one needs to disentangle the RM caused by the Milky-Way, intergalactic medium, and, possibly, the intervening clusters. Observing of RM against gravitational lens systems is a powerful tool to study the magnetic field strength in distant lens galaxies since it is free from contributions of the line-of-sight interveners. The difference of Faraday rotations of a lens system is attributed mainly to the conditions of the magneto-ionic medium of a lens galaxy. The Faraday rotation measurements of lens systems can effectively probe the existence of large-scale regular fields in distant star-forming and elliptical galaxies (Narasimha \& Chitre  \cite{narasimha04}). The SKA with the sensitivity down to an rms of
1~$\mu$Jy will be able to find few thousands of lens systems per square degree with Einstein radii between $0.5''$ and $2''$ typical of galaxy lensing (Koopmans et al.  \cite{koopmans04}). Most of these sources 
are expected to be star-forming galaxies and radio-quiet active galactic nuclei, thus, allowing to probe the cosmological evolution of magnetic fields in these galaxies beyond $z>1$.
In addition, polarized background sources that are lensed into extended arcs can be used, through Faraday rotation, to map RM in the lens itself.

In the hierarchical merger formation model, massive galaxies with
masses greater than the MW and high star-formation rate
($>100$\,M$_{\odot}$) formed at $z\sim$\,(2 to 3). On the other hand,
observations of undisturbed massive galaxies at earlier epochs
support the alternative idea of their formation by accretion of
narrow streams of cold gas, as evident from cosmological simulations
(Dekel et al. \cite{dekel09}). Polarization observations of massive distant
galaxies with the SKA will be crucial to distinguish between
different cosmological scenarios of formation and evolution of galaxies. 

%Another powerful tool to study the magnetic field strength in distant galaxies is the observing of RM against gravitational lens systems. The difference of Faraday rotations of a lens system is attributed mainly to the conditions of the magneto-ionic medium of a lens galaxy, and it is free of intrinsic Faraday rotation of a lens system, Faraday rotation caused by the Milky-Way and intergalactic medium. The Faraday rotation measurements of lens systems can effectively probe the existence of large-scale regular fields in distant star-forming and elliptical galaxies (Narasimha \& Chitre  \cite{narasimha04}). The SKA with the sensitivity down to an rms of
%1~$\mu$Jy will be able to find few thousands of lens systems per square degree with Einstein radii between $0.5''$ and $2''$ typical of galaxy lensing (Koopmans et al.  \cite{koopmans04}). Most of these sources 
%are expected to be star-forming galaxies and radio-quiet active galactic nuclei, thus, allowing to probe the cosmological evolution of magnetic fields in these galaxies beyond $z>1$.
%In addition, polarized background sources that are lensed into extended arcs can be used, through Faraday rotation, to map RM in the lens itself. 

\begin{acknowledgements}
TGA acknowledges support from the DFG-SPP project under grant 566960. This work is also supported by the DFG-RFBR project under grant 08-02-92881.
\end{acknowledgements}

\end{document}